\documentclass[10pt,a4paper]{article}
\pdfoutput=1
\usepackage{jcappub}
\usepackage{graphicx}

\usepackage{subcaption}
\usepackage{amsmath,amssymb}
\usepackage{verbatim}
\usepackage{csquotes}
\usepackage{booktabs}
\usepackage{url}

\newcommand{\apjl}{ApJL}
\newcommand{\apjs}{ApJS}
\newcommand{\aap}{A\&A}

\def\araa{\ref@jnl{ARA\&A}}
\usepackage{xcolor}
\definecolor{notecolor}{rgb}{0.8,0,0}
\definecolor{green}{rgb}{0.0, 0.5, 0.0}

\newcommand{\limpy}{\texttt{LIMpy}}

\usepackage{aas_macros}

\title{Semi-Empirical Approach to [CII] Line Intensity Mapping}

\author[a,b]{Anirban Roy,}
\author[c,d,e,f]{and Andrea Lapi}

\affiliation[a]{Department of Physics, New York University, 726 Broadway, New York, NY, 10003, USA}
\affiliation[b]{Center for Computational Astrophysics, Flatiron Institute, New York, NY 10010, USA}
\affiliation[c]{SISSA, Via Bonomea 265, 34136 Trieste, Italy}
\affiliation[d]{Institute for Fundamental Physics of the Universe (IFPU), Via Beirut 2, 34014 Trieste, Italy}
\affiliation[e]{INFN-Sezione di Trieste, via Valerio 2, 34127 Trieste, Italy}
\affiliation[f]{IRA-INAF, Via Gobetti 101, 40129 Bologna, Italy}

\emailAdd{ar8816@nyu.edu, lapi@sissa.it}

\abstract{The line intensity mapping technique involves measuring the cumulative emission from specific spectral lines emitted by galaxies and intergalactic gas. This method provides a way to study the matter distribution and the evolution of large-scale structures throughout the history of the Universe. However, modeling intensity mapping from ab-initio approaches can be challenging due to significant astrophysical uncertainties and noticeable degeneracies among astrophysical and cosmological parameters. To address these challenges, we develop a semi-empirical, data-driven framework for galaxy evolution, which features a minimal set of assumptions and parameters gauged on observations. By integrating this with stellar evolution and radiative transfer prescriptions for line emissions, we derive the cosmic [CII] intensity over an extended redshift range $0 \lesssim z \lesssim 10$. Our approach is quite general and can be easily applied to other key lines used in intensity mapping studies, such as [OIII] and the CO ladder. We then evaluate the detectability of the [CII] power spectra using current and forthcoming observational facilities. Our findings offer critical insights into the feasibility and potential contributions of intensity mapping for probing the large-scale structure of the Universe and understanding galaxy evolution.}

\keywords{Intensity mapping -- galaxy evolution -- structure formation}

\arxivnumber{---}

\begin{document}

\maketitle
\flushbottom

\section{Introduction}
\label{sec:intro}

Exploring the Universe through redshifted atomic and molecular line emission from galaxies and the intergalactic medium offers a window into the density fluctuations of dark matter and the physics of galaxy formation and evolution. Factors such as star formation history, metallicity, and the halo mass of host galaxies significantly influence the intensity of spectral lines. At high redshifts, the challenge lies in resolving individual galaxies within survey fields, hindering a comprehensive understanding of galaxy formation and their interactions with the intergalactic medium. Several observational efforts have been made to detect the 21 cm line emission to study the cosmic dawn, the epoch of reionization (EoR), and late-time structure formation. Intensity mapping of other lines, such as fine-structure emission from carbon [CII] line ($157.7\, \mu$m), doubly ionized oxygen [OIII] ($88.4\,\mu$m), and rotational emission lines from CO are of particular interest for line intensity mapping (LIM) experiments \citep{Suginohara1998, Righi2008b, Lidz2011_CO, Carilli2011, Fonseca:2016, Gong2017, Kovetz2017LIM_report, Chung2018CII, PadmanabhanCO, Padmanabhan_CII, Dumitru2018, Chung2018CO, Kannan:2021ucy, Murmu:2021ljb, Karoumpis2021, Roy:2024kzc}. Multi-line intensity mapping consolidates emissions from both prominent and faint sources, yielding invaluable insights into galaxy clustering, star formation rate density, and luminosity functions \citep{Visbal2010, Visbal2011, Kovetz2017LIM_report, Bernal:2022jap}. By detecting varied atomic and molecular line intensities across specific observational frequencies, LIM enables the reconstruction of a three-dimensional map of the Universe at different epochs or redshifts, thereby offering a unique perspective on the evolution of cosmic structures.

Several experiments, including COMAP \cite{COMAP:2018kem}, CONCERTO \cite{CONCERTO-science-2020}, EXCLAIM \cite{EXCLAIM-2020}, FYST \cite{CCAT-prime2021}, SPHEREx \cite{SPHEREx-science-paper2018}, SPT-SLIM \cite{SPT-SLIM}, TIME \cite{Time-science-2014}, and TIM \cite{TIM2020}, are currently under development to achieve diverse scientific goals through the observation of LIM. These objectives range from detecting signatures of the EoR to investigating star formation processes within galaxies. Each experiment offers unique frequency coverage, enabling the observation of various atomic and molecular lines that emit at different frequencies. By combining datasets from these experiments, it becomes feasible to perform cross-correlation studies \cite{Schaan:2021gzb, Roy-Lim-LLX}. This approach not only enhances the statistical robustness of the observations but also facilitates validation across a broad range of redshifts to explore the intricate physics underlying galaxy formation and evolution. 

Simulations and analytic models of spectral line emissions reveal significant astrophysical uncertainties in determining their luminosity across redshifts $z\sim 0 - 10$). These uncertainties arise from complex factors such as the availability of cold molecular gas, gravitational instability of gas clouds, feedback from stellar processes like supernovae and stellar winds, and the dynamics of galaxies \cite{Breysse:2021ecm, zhang2023characterizing,Yang2022, Limfast2, zhang2023characterizing}. Environmental conditions, variations in metallicity, and magnetic fields also exert critical influences \cite{Behroozi2019, Garcia:2023djz}. LIM experiments aggregate emissions from multiple galaxies, each with unique interstellar medium properties. These aggregated emissions can be statistically modeled with a mean value and associated scatter \cite{Greve2014}. Traditionally, generating multi-line intensity maps involves assigning line luminosities to halo catalogs derived from $N-$body simulations or semi-numerical approximations \cite{Limfast1, Karoumpis2021, limpy, Niemeyer2023-simple}. This approach requires modeling the function $L_{\rm line}(M_{\rm H})$, which depends on various astrophysical parameters governing line transitions in galaxies. Typically, a simplified model based on empirical scaling relations is used to map the relationship between halo mass and line luminosity. A key assumption is the strong correlation between multi-line luminosities and the star formation rate (SFR). The relationship between SFR and halo mass can be informed by sophisticated simulations of galaxy formation such as IllustrisTNG and EAGLE \cite{TNG-gen, Nelson:2018uso, Eagle-sim-2015}. The scatter in the SFR-to-halo mass relation introduces uncertainty into modeling $L_{\rm line}(M_{\rm H})$.
At lower redshifts, these relationships can often be approximated as power laws in logarithmic scales, but their behavior across a broader redshift range remains uncertain. Nevertheless, such models are widely used in the literature due to their computational efficiency in generating LIM maps and their ability to explore the potential of current and future LIM experiments. Therefore, incorporating these models to derive $L_{\rm line}(M_{\rm H})$, considering the uncertainties between $M_{\rm H}$ to SFR and SFR to $L_{\rm line}$, introduces uncertainties in the power spectrum by up to two orders of magnitude \cite{Karoumpis2021, limpy, clarke2024cii}.

All of these factors above lead to a basic question: how can we interpret LIM observations within the context of parameter estimation to gain a better understanding of astrophysical processes?
In this vein, we take a semi-empirical, data-driven route to address the problem. Semi-empirical models \cite{Aversa2015,Moster2018,Behroozi2019,Grylls2019,Hearin2022,Drakos2022,Fu2022,Boco2023,Zhang2023}  have recently emerged as useful frameworks that constitute a kind of `effective' approach to galaxy formation and evolution, in that they do not attempt to model the small-scale physics regulating the baryon cycle from first principles, but marginalize over it by exploiting empirical relations between the large-scale properties of galaxies. The value of these models stands in that they feature a minimal set of assumptions and parameters gauged on observations.
Semi-empirical approaches are particularly useful when galaxy formation recipes must be coupled with those from other branches of astrophysics (e.g., stellar evolution and radiative transfer models of line emissions in the present context), and therefore it is worth minimizing the uncertainties/hypotheses at least on the former side by exploiting basic data-driven inputs.
For example, LIM observations offer significant potential for extracting both cosmological and astrophysical insights. Modeling their statistical properties, such as the power spectrum, depends crucially on cosmological parameters such as the Hubble expansion rate, and astrophysical quantities like densities of baryonic and dark matter, clustering properties, and halo bias. Achieving precise measurements of these parameters requires minimizing astrophysical uncertainties, necessitating the integration of data from complementary observational probes. Thus the semi-empirical approach may become essential to avoid or at least limit the inherent degeneracies between astrophysical and cosmological parameters in LIM studies.

The plan of the paper is straightforward: in Section \ref{sec|SEM} we present a semi-empirical, data-driven galaxy evolution framework suited for LIM studies. In Section \ref{sec:line_sims} we outline our simulation-based approach for generating LIM maps from our semi-empirical model. In Section  \ref{sec:forecast} we forecast the detectability of these signals. Finally, we summarize our findings and conclusions in Section \ref{sec:discussion}. Throughout this study, we adopt the cosmological model of a flat $\Lambda$CDM Universe, constrained by the cosmological parameters derived from the Planck TT, TE, EE+lowE+lensing analysis \citep{P18:main}. Stellar masses and luminosities of galaxies are derived assuming a Chabrier's \cite{Chabrier2003} initial mass function. A value $Z_\odot\approx 0.0153$ for solar metallicity is adopted.

\section{Semi-empirical model for LIM}\label{sec|SEM}

\begin{figure*}
  \begin{center}
    \includegraphics[width=0.55\textwidth]{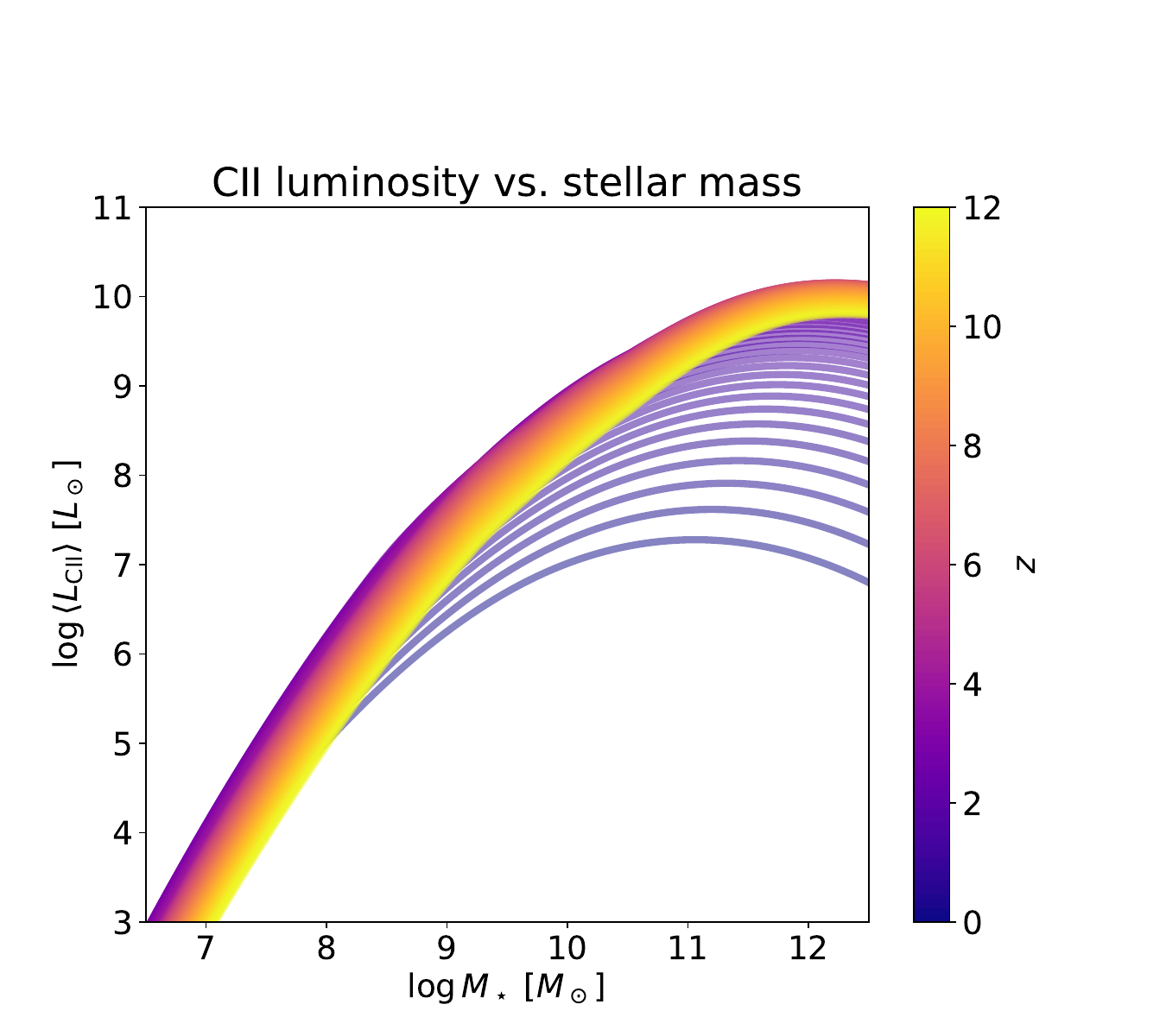}
    \caption{The average [CII] luminosity $L_{\rm [CII]}$ as a function of the stellar mass $M_\star$ at different redshifts (color-coded). This incorporates the distribution of SFR according to the main sequence and of metallicity according to the fundamental metallicity relationship, as detailed in Section \ref{sec|SEM}.}
    \label{fig|L[CII]}
  \end{center}
\end{figure*}

\begin{figure*}
  \begin{center}
    \includegraphics[width=0.55\textwidth]{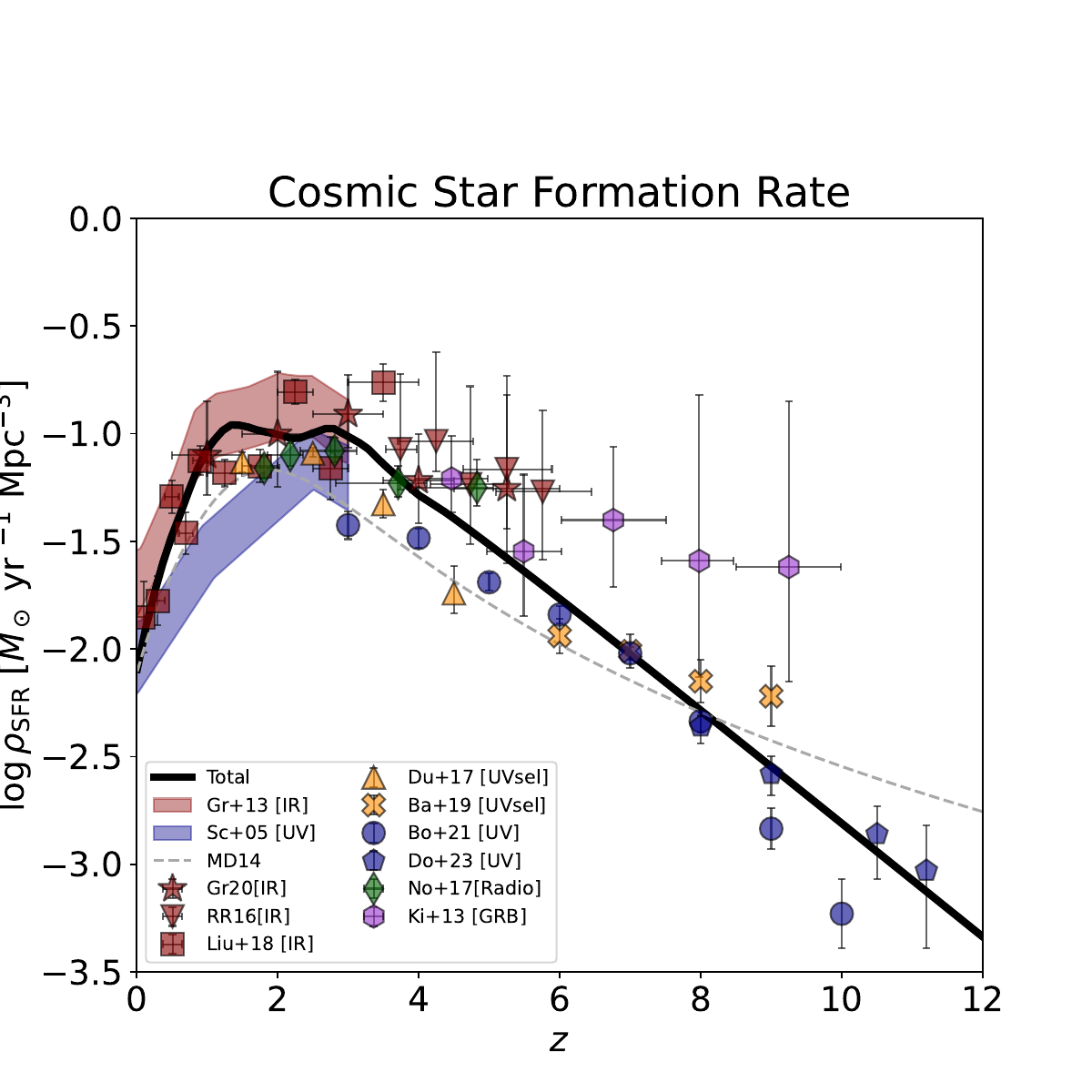}
    \caption{The cosmic star formation rate density as a function of redshift. The outcome of our semi-empirical model is illustrated by the black solid line. Data are from \cite{Gruppioni2013} (blue shaded area), \cite{Schiminovich2005} (blue shaded area), \cite{Gruppioni2020} (red stars), \cite{Rowan2016} (red inverse triangles), \cite{Liu2018} (red squares), \cite{Dunlop2017} (yellow triangles), \cite{Bhatawdekar2019} (yellow crosses), \cite{Oesch2018,Bouwens2021} (blue circles), \cite{Donnan2023} (blue pentagons), \cite{Novak2017} (green rhomboids), and \cite{Kistler2013} (magenta hexagons; from GRB). For reference, the dashed grey line shows the classic determination by \cite{Madau2014rev} based on superseded data at $z\lesssim 4$.}
    \label{fig|SFRD}
  \end{center}
\end{figure*}

We devise a semi-empirical framework suited for LIM studies by combining data-driven quantities related to galaxy formation and evolution with standard recipes for line emissions. For the sake of definiteness, in this Section, we take [CII] emission as a reference, but the same formalism applies to any other line such as [OIII] or the CO ladder.

We start by writing an expression for the average [CII] luminosity in a galaxy with a given stellar mass and redshift:
\begin{equation}\label{eq|L[CII]ave}
\langle L_{\rm [CII]}\rangle(M_\star,z) =  \int{\rm d}\log\psi\, \frac{{\rm d}p}{{\rm d}\log \psi}(\psi|M_\star,z)\,\int{\rm d}\log Z\, \frac{{\rm d}p}{{\rm d}\log Z}(Z|M_\star,\psi,z)\, L_{\rm [CII]}(\psi,Z)~
\end{equation}
where $L_{\rm [CII]}(\psi,Z)$ is the [CII] luminosity at given SFR $\psi$ and metallicity $Z$ from radiative transfer models, ${\rm d}p/{\rm d}\log \psi$ is the observed distribution of SFR at a given stellar mass $M_\star$ and redshift $z$, and ${\rm d}p/{\rm d}\log Z$ is the observed distribution of metallicity at a given stellar mass, SFR and redshift.  As for $L_{\rm [CII]}(\psi,Z)$ we exploit the results of radiative transfer models by \cite{Vallini2015,Lagache2018}, which envisage the [CII] emission to be mainly originated in photo-dissociation regions; their outcomes are well fitted by:
\begin{equation}\label{eq|L[CII]}
\log \frac{\langle L_{\rm [CII]}\rangle}{L_\odot}\approx 7.0+1.2\,\log \frac{\psi}{M_\odot\, {\rm yr}^{-1}} + 0.021 \log \frac{Z}{Z_\odot}+0.012\, \log \frac{\psi}{M_\odot\, {\rm yr}^{-1}}\, \log \frac{Z}{Z_\odot}-0.74\, \log^2 \frac{Z}{Z_\odot}~.
\end{equation}
As to ${\rm d}p/{\rm d}\log \psi$, we exploit the observational finding that at any given cosmic time the distribution of galaxies in the $\psi$ vs. $M_\star$ plane can be represented as a double log-normal distribution around the galaxy main sequence (MS) and the starburst (SB) population. Therefore one can take
\begin{equation}\label{eq|probsfr}
\frac{{\rm d}p}{{\rm d}\log \psi}(\psi|M_\star,z) = \frac{\mathcal{N}_{\rm MS}}{\sqrt{2\,\pi\,\sigma_{\rm MS}^2}}\, e^{-[\log\psi-\log\psi_{\rm MS}]^2/2\sigma_{MS}^2} + \frac{\mathcal{N}_{\rm SB}}{\sqrt{2\,\pi\,\sigma_{\rm SB}^2}}\, e^{-[\log\psi-\log\psi_{\rm SB}]^2/2\sigma_{SB}^2}~,
\end{equation}
with $\mathcal{N}_{\rm MS}=1-\mathcal{N}_{\rm SB}$ being the fraction of galaxies in the MS, and $\mathcal{N}_{\rm SB}(M_\star,z)$ being the fraction of starburst galaxies by \cite{Chruslinska2021}; moreover, $\log \psi(M_\star,z)$ is the latest determination of the MS by \cite{Popesso2023}, and $\log\psi_{\rm SB}=\log_{\rm MS}+0.6$ is the typical location of the SB sequence, while $\sigma_{\rm MS}\approx 0.2$ dex is the dispersion of the MS and $\sigma_{\rm SB}\approx 0.25$ dex that of the SB population.
As to ${\rm d}p/{\rm d}\log Z$, observations indicate that it can be described as a log-normal distribution around the fundamental metallicity relation (FMR), i.e.
\begin{equation}\label{eq|probzeta}
\frac{{\rm d}p}{{\rm d}\log Z}(Z|M_\star,\psi,z) = \frac{1}{\sqrt{2\,\pi\,\sigma_{\rm FMR}^2}}\, e^{-[\log Z-\log Z_{\rm FMR}]^2/2\sigma_{\rm FMR}^2} 
\end{equation}
where $\log Z(M_\star,\psi, z)$ is the average FMR by \cite{Curti2023} including the high-$z$ correction by \cite{Nakajima2023}, and $\sigma_{\rm FMR}\approx 0.15$ dex is its dispersion.

\begin{figure*}
  \begin{center}
    \includegraphics[width=0.55\textwidth]{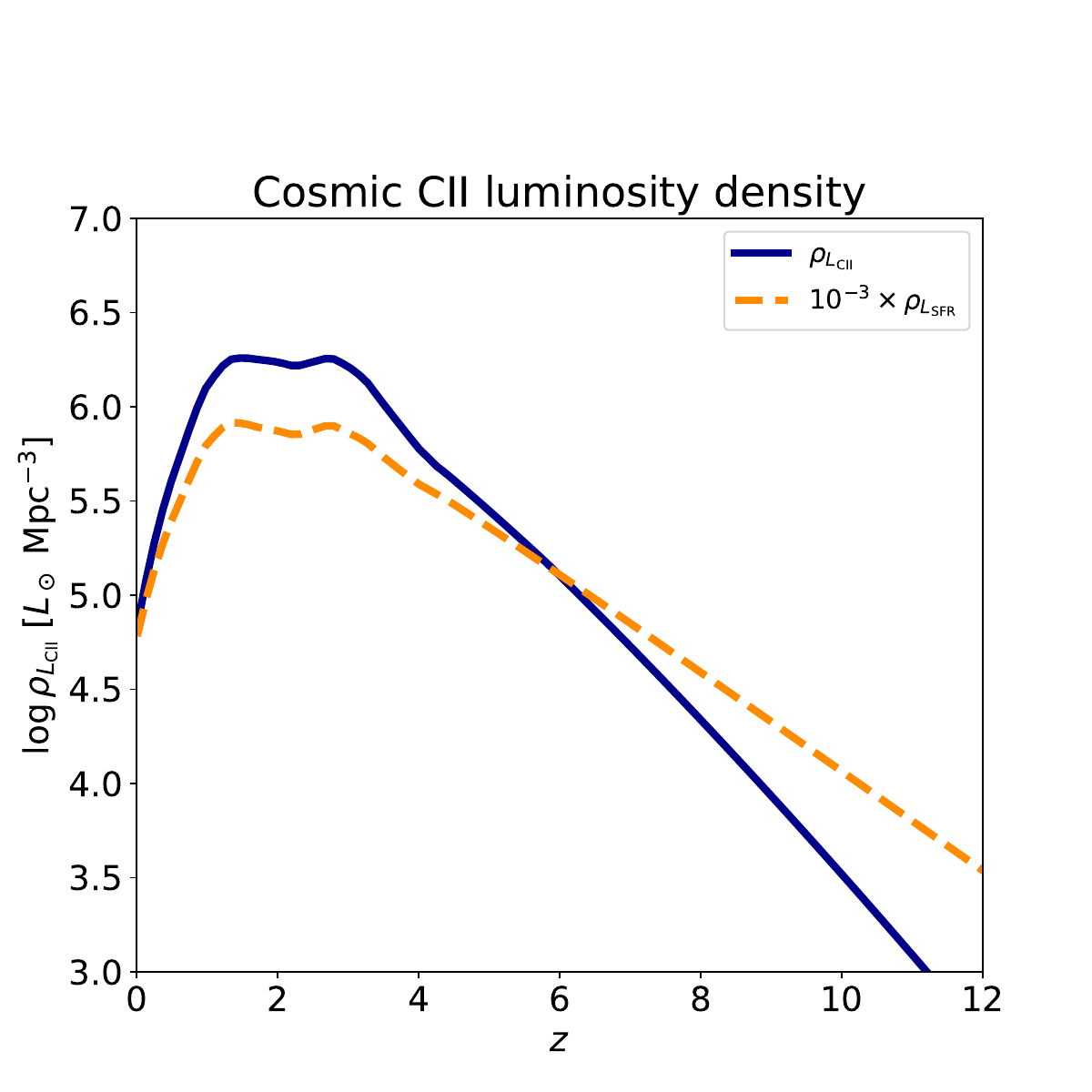}\\
    \includegraphics[width=0.495\textwidth]{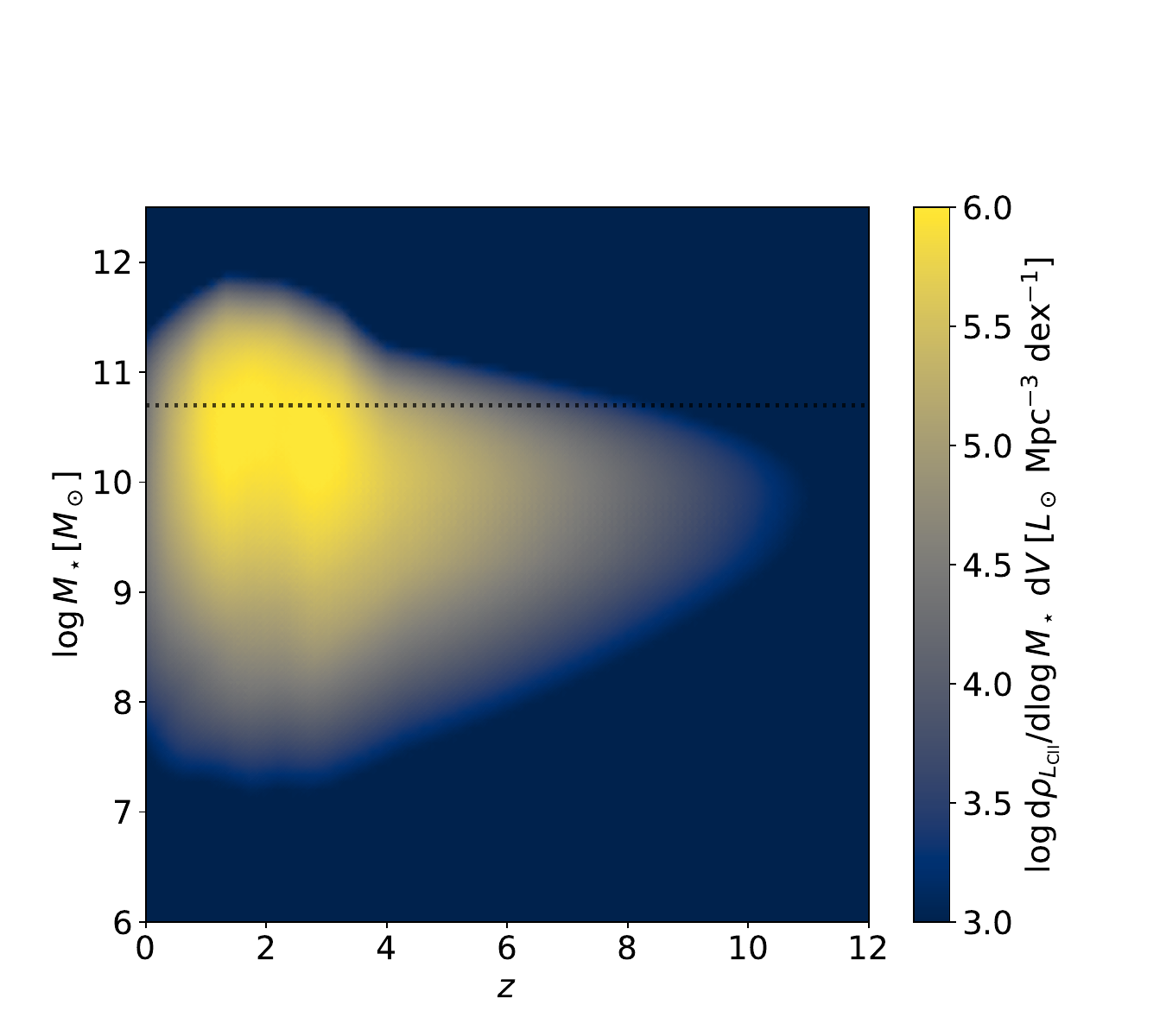}
    \includegraphics[width=0.495\textwidth]{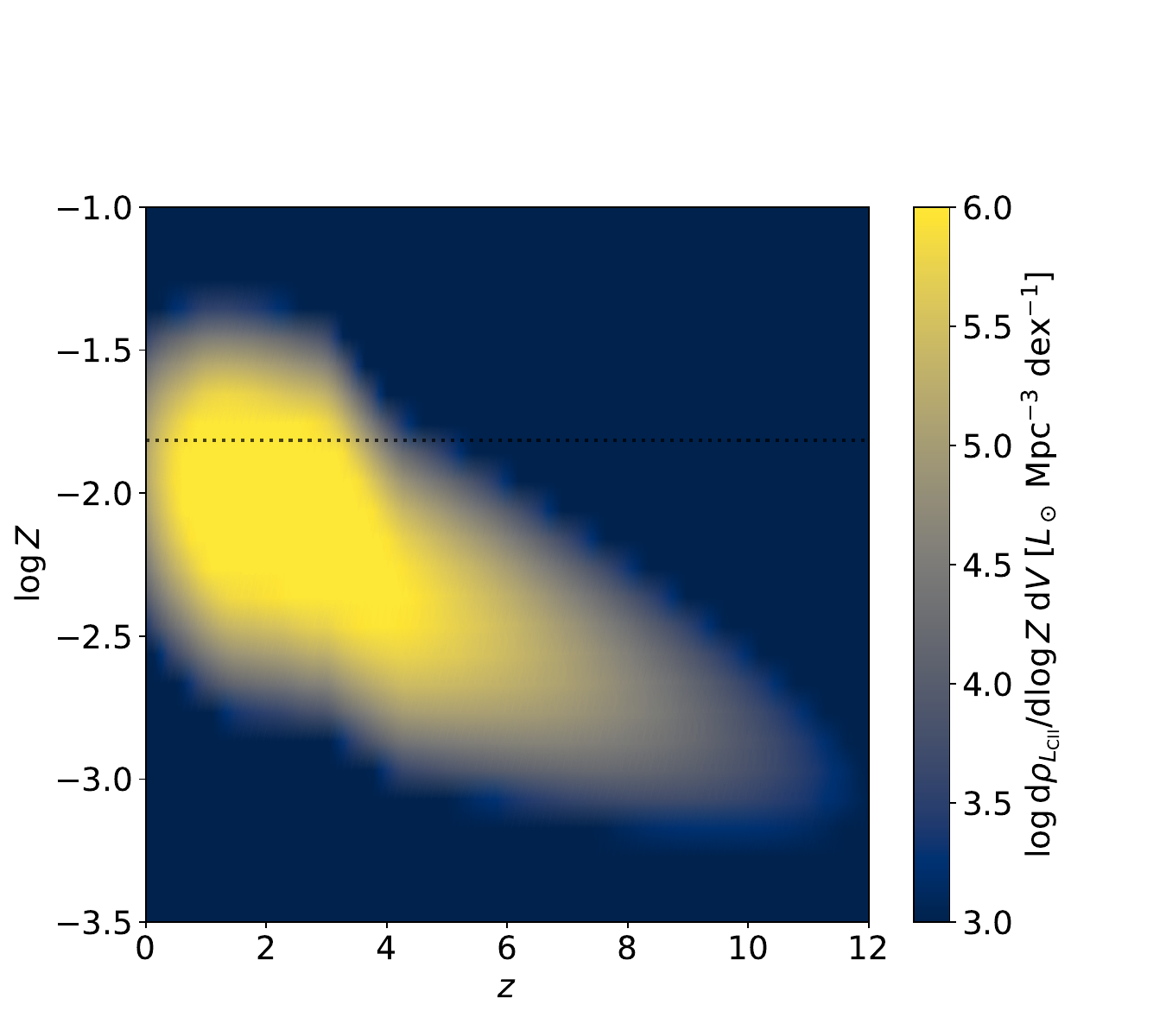}
        \caption{Top panel: the cosmic [CII] luminosity density as a function of redshift; for reference, the orange dashed line displays the luminosity density in SFR scaled down by a factor $10^3$. Bottom panels: the cosmic [CII] luminosity density (color-coded) sliced in stellar mass (left) and metallicity (right) as a function of redshift.}
    \label{fig|rho[CII]}
  \end{center}
\end{figure*}

In Figure \ref{fig|L[CII]} we show the resulting average [CII] luminosity $\langle L_{\rm [CII]}\rangle$ as a function of stellar mass $M_\star$ and redshift $z$. At given redshift, $\langle L_{\rm [CII]}\rangle$ tends to increase with $M_\star$, since massive galaxies are generally more star-forming and more metal-rich; however, especially for low $z\lesssim 1$, at the highest masses $M_\star\gtrsim$ some $10^{11}\, M_\odot$ the [CII] luminosity $\langle L_{\rm [CII]}\rangle$ saturates or even declines somewhat, as a consequence of the shape of the underlying main sequence relation. Such a bending in the SFR vs. stellar mass relation is usually interpreted in terms of a decreasing star-formation efficiency in the most massive galaxies due to the lower availability of cold gas from inflows or feedback processes from nuclear activity. At given stellar mass $M_\star$, the [CII] luminosity $\langle L_{\rm [CII]}\rangle$ increases with redshift up to $z\approx 3$ mainly because galaxies tend to feature higher star formation efficiency, as envisaged by the main sequence; for higher redshifts a mild decline sets in as a consequence of the reduced amount of metals available in galaxies, as prescribed by the fundamental metallicity relation.

\begin{figure*}
  \begin{center}
    \includegraphics[width=0.495\textwidth]{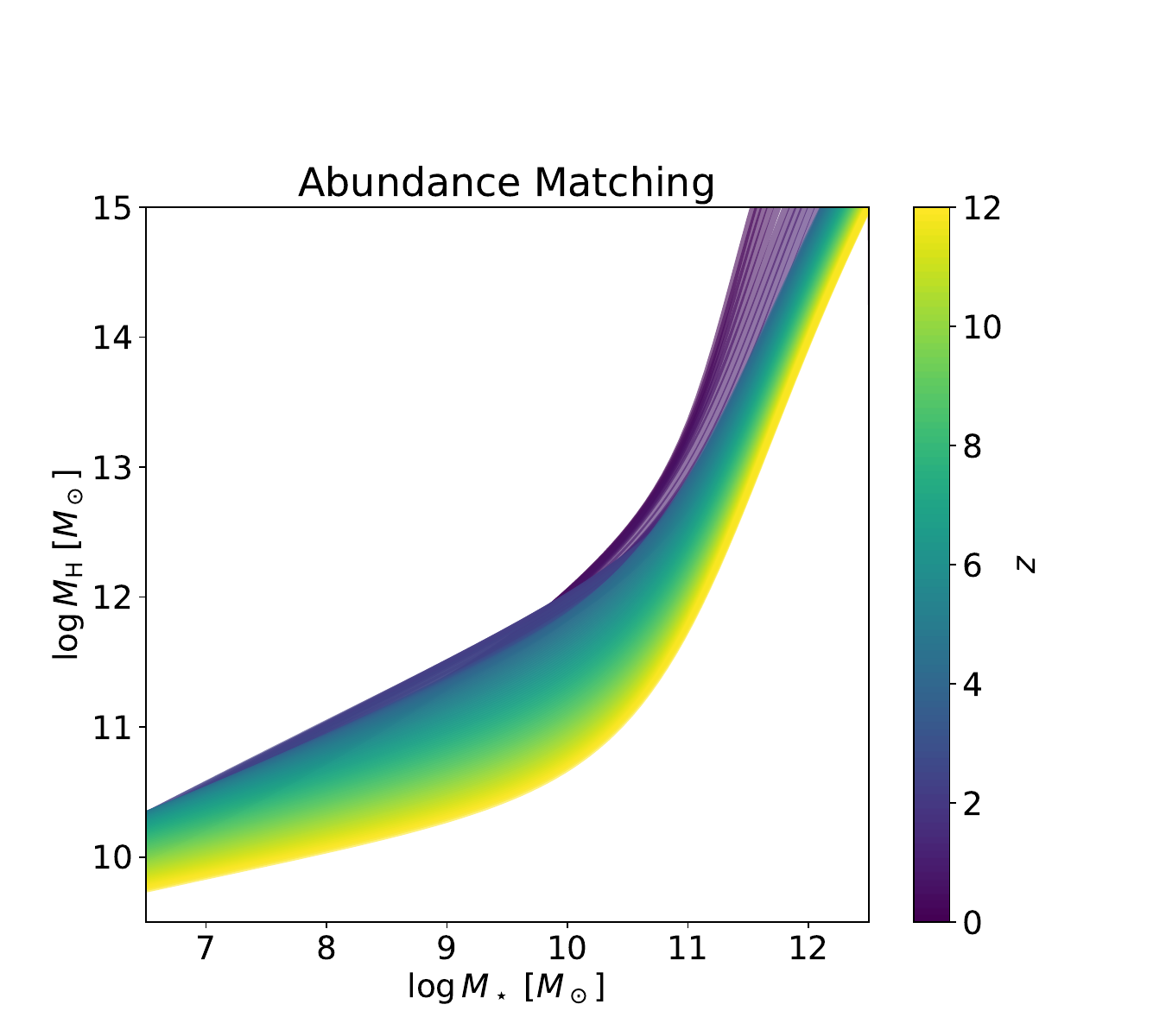}
    \includegraphics[width=0.495\textwidth]{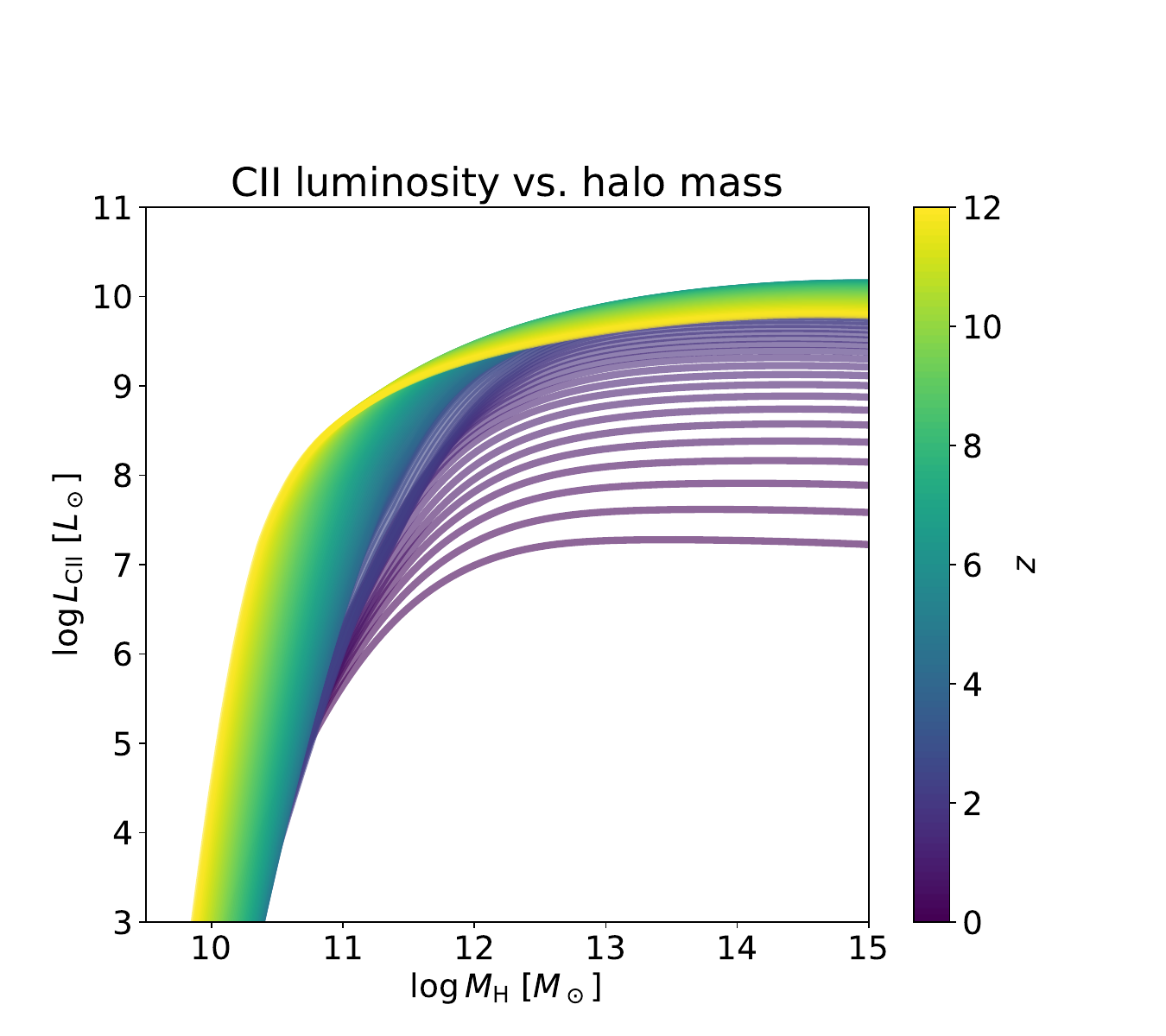}
    \caption{Left panel: the halo vs. stellar mass relationship derived from abundance matching of the stellar mass and halo mass functions, as a function of redshift (color-coded). Right panel: the average [CII] luminosity from Fig.\,\ref{fig|L[CII]} is plotted vs. halo mass at different redshifts (color-coded) via the abundance matching stellar mass-halo mass relation.}
    \label{fig|abma}
  \end{center}
\end{figure*}

The luminosity density in [CII] can be simply computed from Eq. (\ref{eq|L[CII]ave}) via
\begin{equation}\label{eq|rhoL[CII]}
\rho_{L_{\rm [CII]}}(z)= \int{\rm d}\log M_\star\, \frac{{\rm d}N}{{\rm d}\log M_\star\, {\rm d}V}\,\langle L_{\rm [CII]}\rangle(M_\star,z)~,
\end{equation}
where the quantity ${\rm d}N/{\rm d}\log M_\star\, {\rm d}V$ is the latest observational determination of the redshift-dependent galaxy stellar mass function for star-forming galaxies by \cite{Weaver2023}. Incidentally, note that substituting $\langle L_{\rm [CII]}\rangle(M_\star,z)$ with just the SFR $\psi$ or the associated total luminosity $L_{\rm SFR}\approx 10^{10}\, L_\odot\, (\psi/M_\odot\,{\rm yr}^{-1})$ will originate the SFR luminosity density, namely the celebrated `Madau' plot, that can be exploited as a preliminary validation of our semi-empirical approach. The outcome is shown in Figure \ref{fig|SFRD}, and well agrees with a wealth of observational estimates from data selected in various electromagnetic bands at different redshifts \cite{Gruppioni2013,Schiminovich2005,Gruppioni2020,Rowan2016,Liu2018,Dunlop2017,Bhatawdekar2019,Oesch2018,Bouwens2021,Donnan2023,Novak2017,Kistler2013}. The double peak around the `cosmic noon' at $z\approx 2$ is not to be taken as a realistic feature since it can be traced back to the detailed shape of the stellar mass function fits by \cite{Weaver2023}. We also stress that stellar mass function determinations are robust out to $z\sim 6$, so at higher redshifts, the outcome should be taken with care, and considered as an educated extrapolation albeit it is found in remarkable accord with the latest determinations by the JWST \cite{Donnan2023}.

In Figure \ref{fig|rho[CII]} we illustrate the [CII] luminosity density as a function of redshift, and compare it to the SFR luminosity density from the previous Figure (scaled down appropriately to ease the comparison). The shapes of the two curves are very similar to the cosmic noon at $z\sim 2$, but for higher redshift the [CII] luminosity density declines appreciably faster. This is mainly due to the dependence on [CII] luminosity on metallicity that implies a reduced [CII] emission in high redshift galaxies that on average are poorer in metals. The trends with stellar mass and metallicity at different redshifts can be perceived better in the lower panels of Figure \ref{fig|rho[CII]}. These show that most of the [CII] luminosity density comes out to the cosmic noon at $z\sim 2-3$ from galaxies with stellar masses a few $10^{10}-10^{11}\, M_\odot$ and metallicities $Z_\odot/3-Z_\odot$. Nevertheless, out to $z\sim 3$ there is an appreciable contribution to $\rho_{L_{\rm [CII]}}$ from galaxies with $M_\star$ down to $10^8\, M_\odot$, while at higher $z$ this minimum mass increases somewhat to around $10^9\, M_\odot$ while the contribution from massive galaxies is also limited given their reduced number density. For the same reason, in moving toward higher redshift the galaxies contributing most to $\rho_{L_{\rm [CII]}}$ feature smaller stellar masses and hence lower metallicities down to $Z\lesssim Z_\odot/10$.

\begin{figure*}
  \begin{center}
    \includegraphics[width=0.55\textwidth]{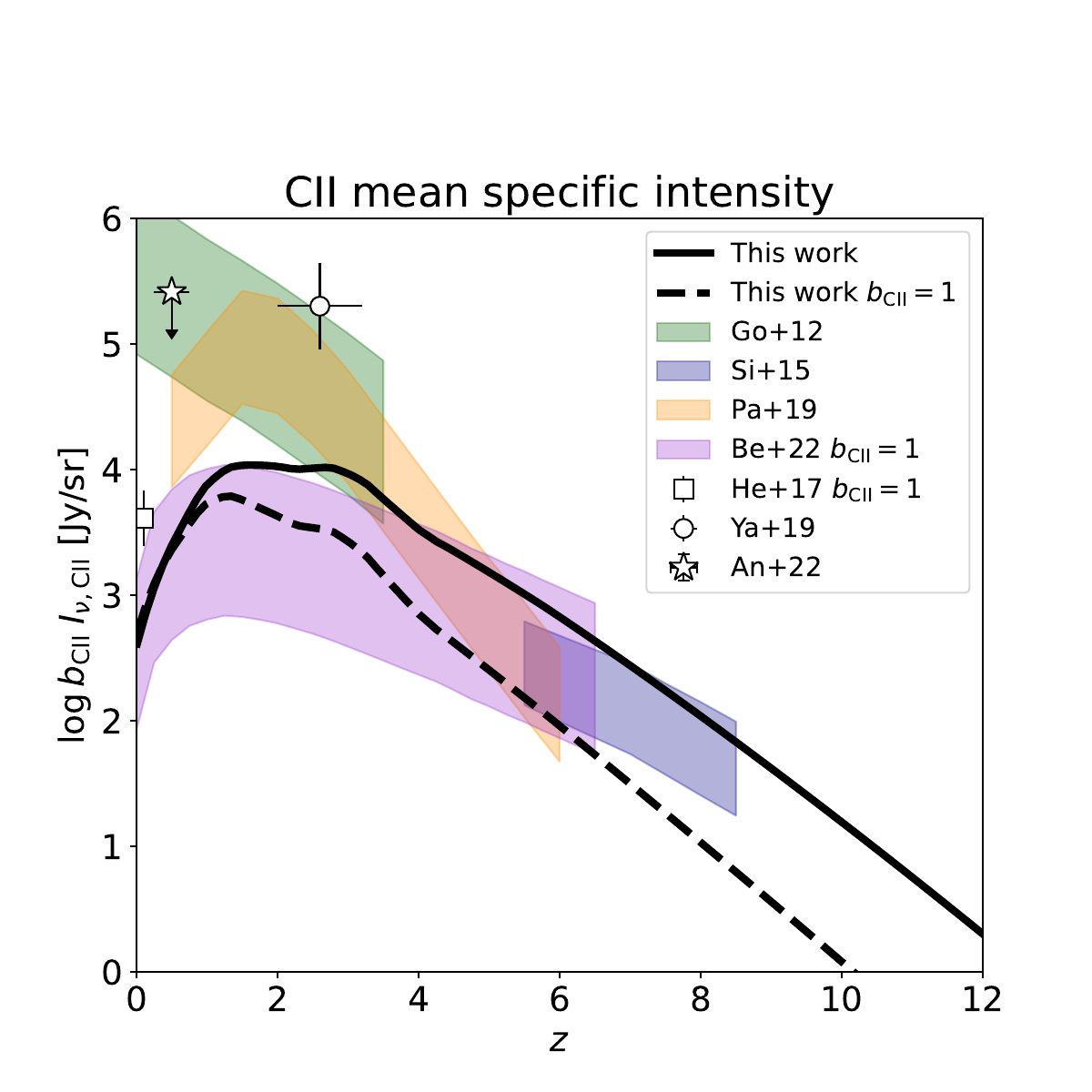}
    \caption{The [CII] mean specific intensity as a function of redshift from our semi-empirical model: solid line refers to the quantity $b_{\rm [CII]}\, I_{\nu,\rm [CII]}$, while dashed line to $I_{\nu, \rm [CII]}$ by assuming $b_{\rm [CII]}=1$. The magenta area reports the uncertainty range from models based on scaling relations between [CII] luminosity and SFR, according to the review by \cite{Bernal2022}. The orange area shows the estimate by \cite{Padmanabhan2019} aimed at reproducing the determination by \cite{Pullen2018}. The green shaded area displays the range of collisional excitation models by \cite{Gong2012}. The blue area illustrates the range covered by the high-redshift models from \cite{Silva2015}. The white squares are the estimate based on the local [CII] luminosity function by \cite{Hemmati2017}, the white circle is from the \emph{Planck}$\times$QSO cross-correlation measurements by \cite{Yang2019}, and the star is an upper limit from the FIRAS$\times$BOSS cross-correlation measurements by \cite{Anderson2022}.}
    \label{fig|Inu[CII]}
  \end{center}
\end{figure*}

\begin{figure*}
  \begin{center}
\includegraphics[width=0.44\textwidth]{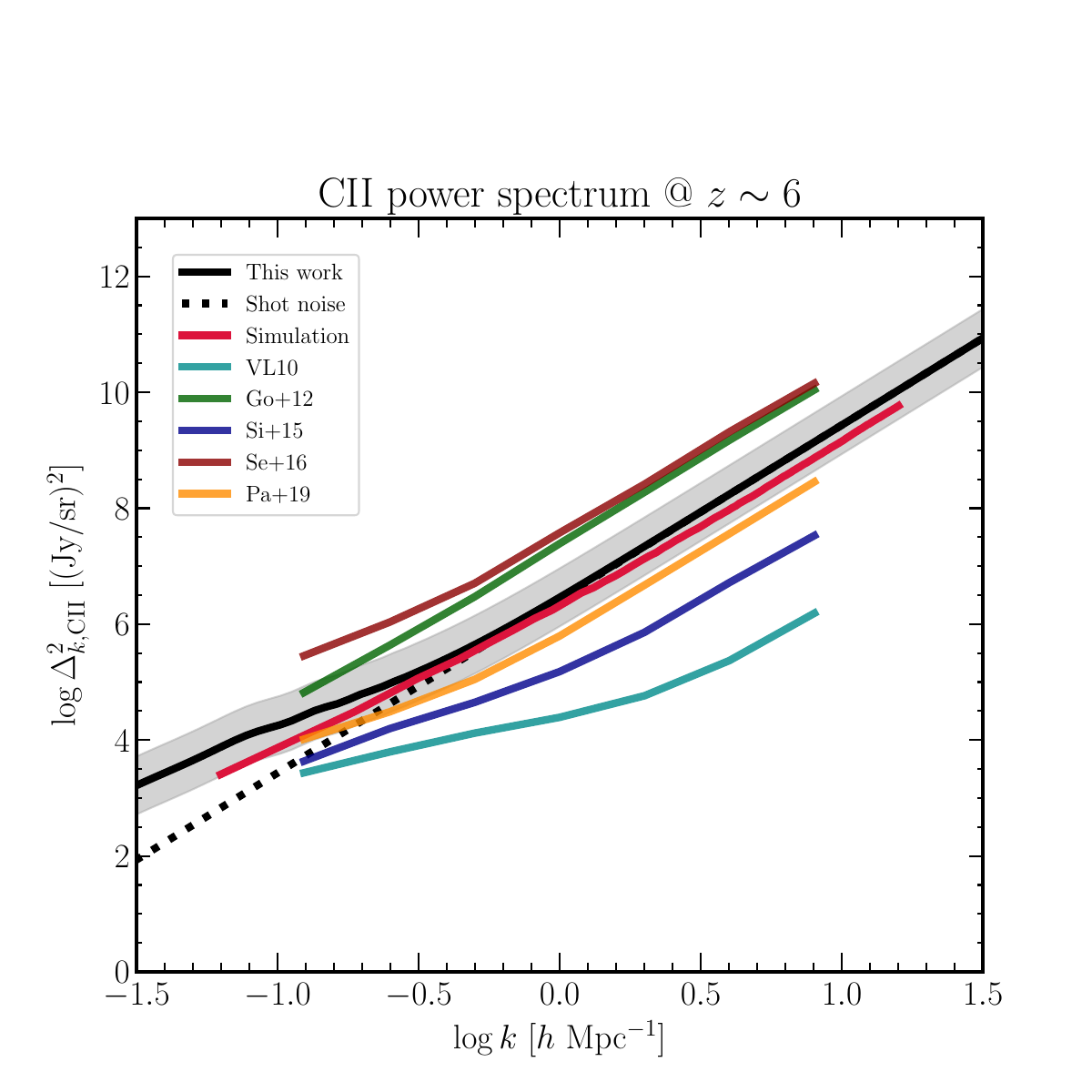}    
\includegraphics[width=0.49\textwidth]{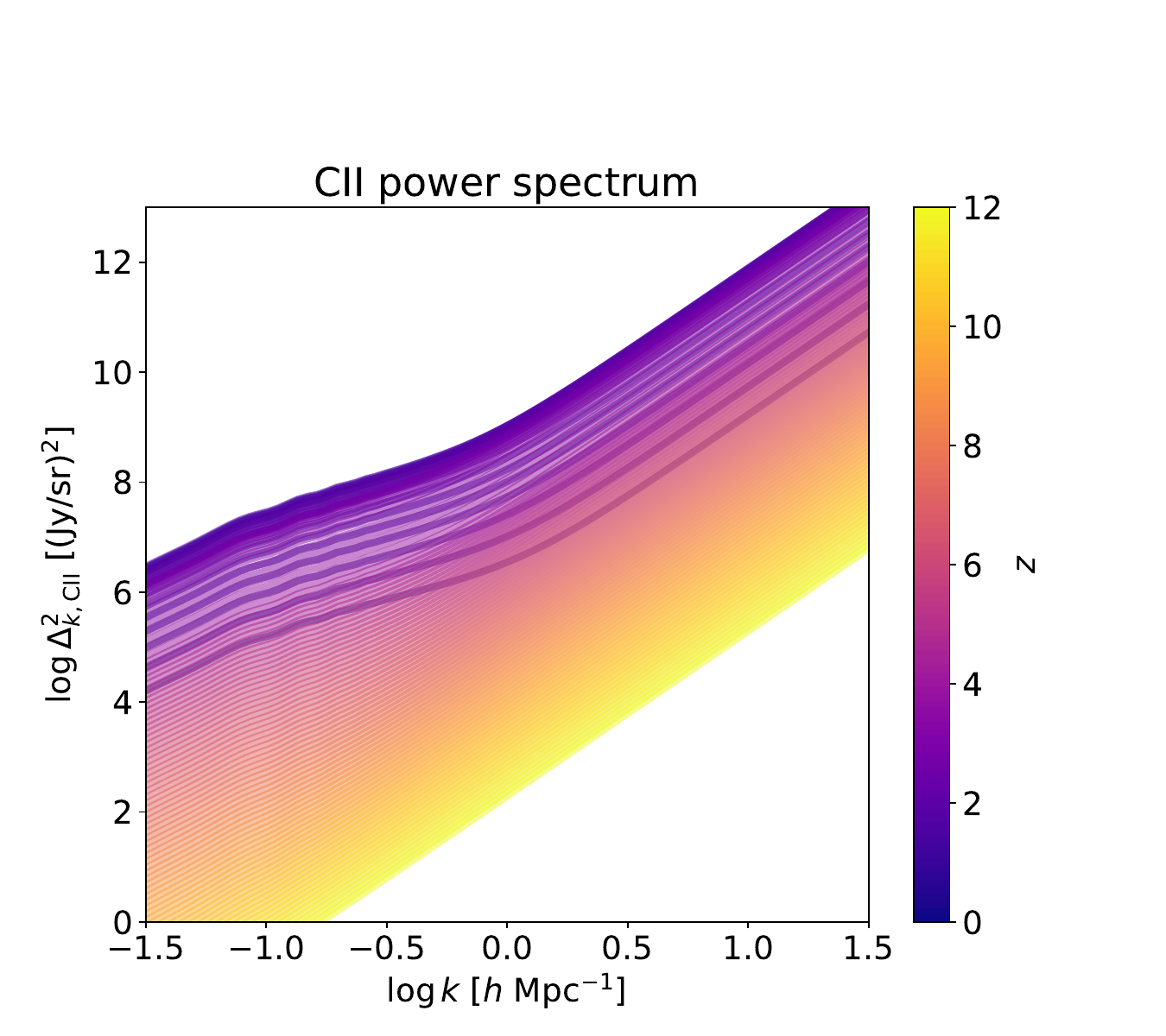}\\
    \caption{Left panel: The [CII] power spectrum at $z\sim 6$ from our semi-empirical model (solid black line: total, dashed line: shot noise contribution, shaded area:  typical uncertainty) is compared to other literature determinations from \cite{Visbal2010} (cyan), \cite{Gong2012} (green), \cite{Silva2015} (blue), \cite{Serra2016} (red), and \cite{Padmanabhan2019} (orange). Right panel: the [CII] power spectrum at different redshifts (color-coded) from our semi-empirical framework.}
    \label{fig|Dk2[CII]_z6comp}
  \end{center}
\end{figure*}

For line intensity mapping, the crucial quantity is the [CII] average specific intensity as a function of redshift, which is just proportional to the above [CII] luminosity density \cite{Padmanabhan2019,Bernal2022}
\begin{equation}\label{eq|Inu[CII]}
I_{\nu,\rm [CII]}(z)\equiv \frac{c}{4\pi}\, \frac{ \rho_{\rm L_{\rm [CII]}}(z)}{\nu_{\rm [CII]}\, H(z)}~,
\end{equation}
where $\nu_{\rm [CII]}\approx 1900$ GHz ($\lambda_{\rm [CII]}\approx 158\, \mu$m) is the rest-frame frequency of the [CII] emission, and $H(z)$ is the Hubble rate. The [CII] power spectrum is then given by
\begin{equation}\label{eq|Pk[CII]}
P_{\rm [CII]}(k,z) = I_{\nu,\rm [CII]}^2(z)\, [b_{\rm [CII]}^2(z)\, P_{\rm lin}(k,z)+P_{\rm shot, [CII]}(k,z)]~,
\end{equation}
where $P_{\rm lin}$ is the linear matter power spectrum, $b_{\rm [CII]}(z)$ is the bias (clustering component) and $P_{\rm shot, [CII]}$ is the shot noise. Specifically, we compute
the shot noise as:
\begin{equation}\label{eq|Pshot}
P_{\rm shot, [CII]}(k,z) = \cfrac{1}{\rho_{L_{\rm [CII]}}^2(z)}\, \int{\rm d}\log M_\star\, \frac{{\rm d}N}{{\rm d}\log M_\star\, {\rm d}V}\,\langle L_{\rm [CII]}\rangle^2(M_\star,z)
~.
\end{equation}
and the clustering component as
\begin{equation}\label{eq|bias}
b_{\rm [CII]}(z) = \cfrac{1}{\rho_{L_{\rm [CII]}}(z)}\, \int{\rm d}\log M_\star\, \frac{{\rm d}N}{{\rm d}\log M_\star\, {\rm d}V}\,\langle L_{\rm [CII]}\rangle(M_\star,z)\, b_{\rm H}(M_{\rm H},z)~,
\end{equation}
where $b_{\rm H}(M_{\rm H},z)$ is the halo bias extracted from the $N$-body simulations by \cite{Tinker2010}. However, the halo bias is given in terms of halo mass $M_{\rm H}$, while our semi-empirical framework provides $\langle L_{\rm [CII]}\rangle(M_\star,z)$
as a function of $M_\star$; therefore to perform the integration in Eq. (\ref{eq|bias}) we need to relate halo to stellar masses. We obtain such relation via abundance matching \cite{Aversa2015} of the stellar and the halo mass functions, i.e., we require the cumulative number densities in galaxies and halos to match as: 
\begin{equation}\label{eq|abma}
\int_{M_{\rm H}}^{+\infty}{\rm d}M_{\rm H}'\;\frac{{\rm d}N}{{\rm d}M_{\rm H}'\, {\rm d}V}(M_{\rm H}',z) = \int^{+\infty}_{M_\star}{\rm d}M_{\star}'\;\frac{{\rm d}N}{{\rm d}M_\star'\, {\rm d}V}(M_\star',z)
\end{equation}
which yields a relation $M_{\rm H}(M_{\star},z)$. The integrand on the left hand side is the halo mass function by \cite{Tinker2008}. We compute the halo mass function, halo bias, and linear matter power spectrum via the \texttt{COLOSSUS} package \cite{Diemer2018}.

The outcome of the abundance matching procedure is displayed in Figure \ref{fig|abma}, in terms of the halo mass vs. stellar mass and of the ensuing [CII] luminosity vs. halo mass relationships. At fixed redshift, the halo mass corresponding to a given stellar mass increases steadily. Specifically, for $M_\star\lesssim 10^{10}\, M_\odot$ the relation is quite flat, meaning that a small change in $M_{\rm H}$ corresponds to a considerable change in $M_\star$: this is the regime where star formation efficiency is thought to be regulated by stellar feedback processes (e.g., cooperative action of stellar winds and supernova explosions). For $M_\star\gtrsim 10^{10}\, M_\odot$ the relation gets steeper, since the star formation efficiency lowers to even smaller values: the physical interpretation calls into play reduced gas inflows, inefficient cooling in large halos, or feedback processes from nuclear activity. In terms of redshift evolution, the typical $M_{\rm H}$ associated with a given $M_\star$ lowers toward high $z$, meaning that the star-formation efficiency increases appreciably: this may be due to more efficient cooling associated to higher gas fraction, enhanced densities, and larger gas clumping present at early times. The $L_{\rm [CII]}$ vs. $M_{\rm H}$ halo mass relation is just a recasting of Figure \ref{fig|L[CII]} in terms of halo mass. It is worth highlighting that there is a strong decline of the [CII] luminosity in halos with mass $M_{\rm H}\lesssim$ a few $10^{10}\, M_\odot$; in terms of our semi-empirical framework, this partially justifies literature approaches where it is assumed that the [CII] emission is strongly suppressed in such low mass halos (e.g., \cite{Padmanabhan_CII}).

In Figure \ref{fig|Inu[CII]} we showcase the [CII] intensity as a function of redshift, either including or not the bias $b_{\rm CII}$, from our semi-empirical framework. The effect of the bias is particularly evident for $z\gtrsim 1$,  contributing to raise the [CII] intensity appreciably (a factor $2-3$) at the cosmic noon $z\sim 2-3$ and considerably (a factor up to $10$) for $z\gtrsim 4$.
Our data-driven estimate of $I_{\rm [CII]}$ is in good agreement with other literature results based on the scaling relation between [CII] luminosity and SFR \cite{Silva2015,Padmanabhan2019,Bernal2022}.
At high redshift $z\gtrsim 4$ it also agrees with collisional excitation models \cite{Gong2012}. Note, however, that though toward low-$z$ the latter models are well known to predict much higher intensities, more in line with cross-correlation constraints \cite{Yang2019,Anderson2022}; this is not surprising since in such models a substantial contribution to [CII] emission at low redshifts comes from hot diffuse gas outside star-forming galaxies, which is plainly not included in our framework or in similar models based on scaling relations. However, for reionization oriented studies this is not a big deal, since as mentioned above all models tend to converge and agree for $z\gtrsim 4$. 

In Figure \ref{fig|Dk2[CII]_z6comp} we show the [CII] power spectrum at different redshifts from our semi-empirical model, and the outcome at $z\sim 6$ is compared with other literature determinations. At scales smaller than $\sim 1\, h^{-1}$ Mpc the power spectrum is mainly contributed by the shot noise, while at smaller scales the clustering term progressively takes over. At fixed scale, the power spectrum follows the redshift evolution of the specific intensity, rising from the present out to $z\approx 2-3$ and then progressively declining toward higher $z$. The outcome from our semi-empirical framework remarkably agrees with other literature estimates \cite{Visbal2010,Gong2012,Silva2015,Serra2016,Padmanabhan_CII}, striking an intermediate course among scaling relation-based approaches and collisional excitation models.

\section{Simulations of line intensity maps}
\label{sec:line_sims}
We have integrated the semi-empirical framework of the previous Section, referred to as \texttt{Roy24}, into the $\limpy$ package \cite{limpy}. This addition is useful because it allows us to compare its predictions with those from other existing models and significantly reduces the computational cost for the generation of intensity maps. The \texttt{Roy24} model does not rely on existing SFR models for creating LIM maps. Instead, it provides a straightforward method for calculating the [CII] luminosity of halos, determining the power spectrum using halo-model prescriptions, simulating maps, and analyzing tools to forecast signal-to-noise ratios (SNRs). Specifically, we performed a two-dimensional interpolation between stellar mass ($M_{\star}$), redshift ($z$), and [CII] luminosity ($L_{\text{CII}}$). We then established the relationship between halo mass and $M_{\star}$ using abundance matching techniques as previously described. This approach enables us to assign [CII] line intensities to halos in the external halo catalogs derived from $N-$body simulations. This method is particularly effective for generating target signal intensity maps and interloper maps, which are crucial for creating mock observations.

To generate the intensity maps, we created halo catalogs by performing $N-$body simulations using the GADGET software package \cite{gadget4}. Our method involved generating $100$ slices spanning a redshift range from $0$ to $20$, with each slice corresponding to an approximate age of $130$ Myr, roughly equivalent to the light travel time across the volume. The dimensions of the simulation box were set to $100$ Mpc/$h$, and we achieved a spatial resolution of approximately $0.156$ Mpc/$h$ using a grid size of $N_{\text{grid}} = 512$. This resolution allowed us to accurately resolve halo masses down to $10^{10} \, M_{\odot}/h$. The same random seed was used to generate snapshots for all $100$ redshifts. This collection of snapshots, all evolving from a single Gaussian random initial condition while maintaining the same seed, is referred to as ``Sim-set 1''. Furthermore, to account for variations along different lines of sight and to capture the intrinsic sample variance, we repeated this process $13$ times, each with a different random seed. As a result, we obtained $13$ distinct sets of simulations (``Sim-set 1'' to ``Sim-set 13''), each evolving from different Gaussian random fields and providing unique representations of the Universe \citep[see][for more details]{Roy-Lim-LLX}. 

\begin{figure*}
  \begin{center}
    \includegraphics[width=\textwidth]{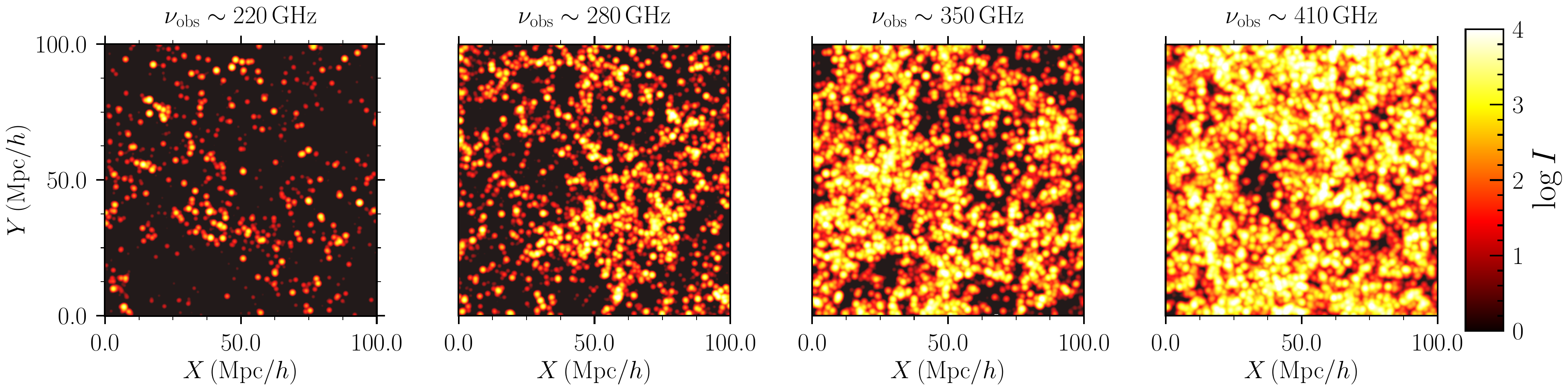}\\
    \caption{We present simulated [CII] intensity maps at four redshifts: $z \approx 7.6$, $5.8$, $4.4$, and $3.6$, corresponding to observational frequencies of $220$ GHz to $410$ GHz. The simulations cover a $100$ Mpc/$h$ box size with 512 grid points along the box axes, with a minimum mass resolution of $10^{10}\,M_\odot/h$. These panels illustrate the signal amplitude, with source clustering increasing at lower redshifts due to non-linear structure formation in the late time Universe.}
    \label{fig:cii_intensity_maps}
    \label{fig|maps}
  \end{center}
\end{figure*}

We present the simulated [CII] intensity maps using the \texttt{Roy24} model in Figure \ref{fig|maps}. Four redshifts have been selected for displaying the intensity maps, corresponding to frequencies ranging from $220$ GHz to $410$ GHz, which aligns with the coverage of FYST's EoR-Spec. These panels illustrate the evolution of structures with respect to the age of the Universe. At high redshifts ($z \gtrsim 6$), high-mass halos have not yet formed, which explains their absence in the simulation box. For example, the most massive halos present in the simulation boxes at $z = 3.6$ and $z = 7.6$ are $1.2 \times 10^{13} \, M_\odot/h$ and $1.2 \times 10^{11} \, M_\odot / h$, respectively. Additionally, the number of galaxies per unit volume is significantly lower at higher compared to lower redshifts, resulting in a fainter intensity map at $z\approx 7.6$ compared to $z \approx 3.6$. Specifically, there are 15.5 times more halos in the lowest redshift box compared to the highest one. We find that the intensity maps at $z \approx 3.6$ are $7$, $26$, and $65$ times brighter than those at $z \approx 4.4$, $5.6$, and $7.6$, respectively.

To analyze the statistical properties, we compute the power spectrum from the simulated [CII] intensity maps using the formula
\begin{equation}
\Delta^2_{\rm [CII]}(k) = \frac{1}{V_{\text{box}}} \frac{k^3}{2\pi^2} \langle \tilde{I}^2(k) \rangle,
\end{equation}
where $\tilde{I}(k)$ denotes the Fourier transform of the voxel intensities and $V_{\text{box}}$ is the volume of the simulation box. This method enables us to analyze the distribution of intensity fluctuations across various spatial scales, providing insight into the spatial characteristics of the [CII] emission within the simulation.

\begin{figure*}
  \begin{center}
\includegraphics[width=\textwidth]{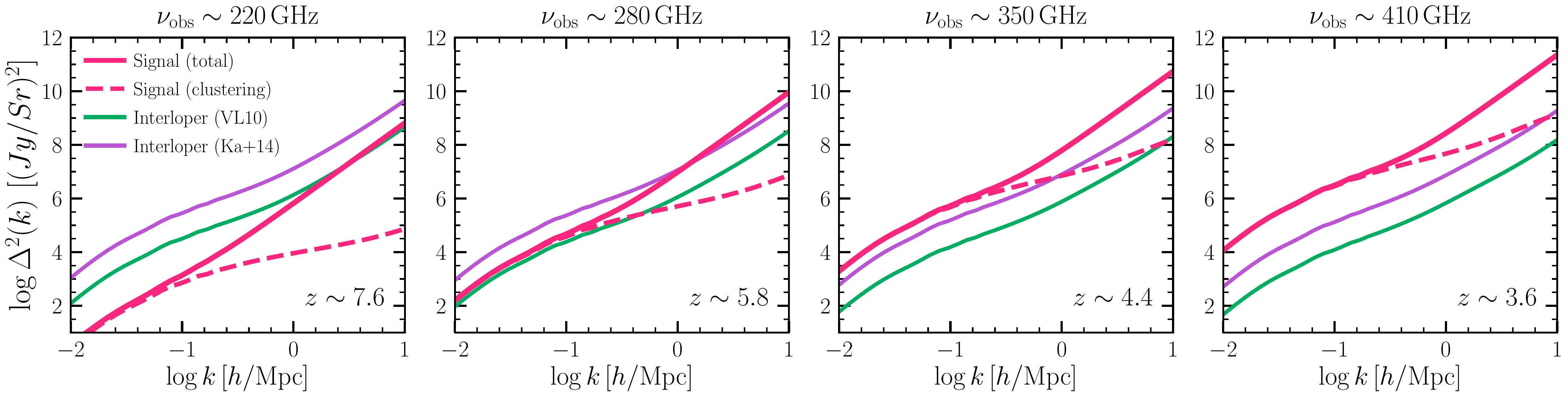}\\
    \caption{We present the [CII] power spectrum as the target signal at four different observational frequencies (solid lines), along with the clustering term (dashed line). Interlopers include all CO J-level transitions for faint lines (VL10; \citep{Visbal2010}) and bright lines (Ka+14; \citep{Kamenetzky2016}). We find that the total signal, comprising clustering and shot noise terms, is dominated by interlopers at $z \sim 7.6$. However, the total signal dominates over interlopers at $z \sim 3.6$ and $4.4$.}
    \label{fig|lim_ps_all}
  \end{center}
\end{figure*}

Figure \ref{fig|Dk2[CII]_z6comp} shows a comparison between the power spectrum derived from the simulation and the power spectrum estimated analytically. The simulation results exhibit a strong agreement with the analytical power spectrum at intermediate scales, specifically in the range $-0.5 \lesssim \log k \lesssim 0.5$. This suggests that our simulation accurately captures the power spectrum's behavior within this range, with the results falling well within the uncertainty bounds of the analytical estimates. However, the differences are primarily attributed to the minimum halo mass considered in the simulations, which is $2 \times 10^{10} \, M_\odot / h$. In contrast, the analytical calculations are based on a minimum mass of $10^{10} \, M_\odot / h$. This variation in minimum mass impacts the power spectrum, highlighting the importance of considering mass thresholds when comparing simulated and analytical results.

\section{Implications for LIM experiments}
\label{sec:forecast}

In this section, we explore the potential for measuring the [CII] signal as estimated from our model, with a focus on both current and future experiments. To model the instrumental noise, we use an example based on FYST's EoR-Spec-like experiment. For all forecasts, we have excluded contributions from atmospheric noise and the complexities arising from the telescope's scanning strategy. We need to account for the fact that lines from different frequencies can be redshifted into a given frequency channel and act as interlopers, creating an obstacle to the detection of high-redshift [CII] emission. Our objective is to forecast the detectability of the [CII] signal while considering both these interlopers and instrumental noise. For each frequency channel, we first identify the redshifts of CO emissions that contribute to the interloper signals, acknowledging that both interlopers and the target signal are frequency-dependent. For instance, in the frequency channel at 220 GHz with a 40 GHz bandwidth, CO lines such as CO(2-1) at $z \approx 0.04$, CO(3-2) at $z \approx 0.57$, CO(4-3) at $z \approx 1.09$, and CO(5-4) at $z \approx 1.61$ can act as contaminants for the target [CII] signal from $z \sim 7.6$. Similarly, for the 410 GHz channel with a 40 GHz bandwidth, CO(4-3) at $z \approx 0.12$, CO(5-4) at $z \approx 0.4$, and CO(6-5) at $z \approx 0.68$ serve as interlopers for the [CII] signal from $z \sim 3.6$.

To model the various CO transitions, we utilize the empirical models available in $\limpy$ \cite{limpy}. Our analysis shows that the models \texttt{Kamenetzky15} and \texttt{VL10} generate bright and faint interlopers, respectively. At $k \sim 0.5\, h/$Mpc, the ratio between the bright and faint interlopers is 9.2 for 220 GHz and 10.7 for 410 GHz. We assume that interloper profiles for other models will fall within this range, which provides a robust basis for our forecasts and ensures that our predictions account for potential variability in interloper signals. To estimate the total power spectrum of the interlopers, we first compute the power spectrum for each individual interloper line at its respective redshift and then sum these spectra to obtain the total contribution from all interlopers.

We compare the target signal with the estimated range of interlopers in Figure \ref{fig|lim_ps_all}. At $\nu_{\text{obs}} = 220 \, \text{GHz}$, the bright interloper is 5.2 times larger and the faint interloper is $47.6$ times larger than the target signal at $k \approx 0.5 \, h/$Mpc. In contrast, for the $410$ GHz frequency channel, the total target signal dominates across all scales, being 26 times larger than the brightest interloper and $287$ times larger than the faintest interloper at the same scale. Where the clustering term dominates the shot noise term, at $k \sim 0.1 \, h/$Mpc, the faintest interloper is $23$ times larger than the signal at $220$ GHz, while the signal is $254$ times larger than the interloper at $410$ GHz.

We adopt the white noise power spectrum for a reference experiment, such as FYST's EoR-Spec, to forecast the detectability of the signal predicted by our model \citep[see Table 1 of][for the noise power spectrum]{CCAT-prime2021}. The error in the measurement of the power spectrum at the $k_i$ bin with width $\Delta k_i$ is estimated as
\begin{align}
    \sigma_{\rm i}(z) = \frac{[P_{\rm [CII]}(k, z) +  P_{\rm int}(k, z)+  P_{\rm WN}(k, z)]^2}{N_m(k_i, \Delta k_i, V_{\rm Surv})}\,.
\end{align}
Here, $P_{\rm [CII]}$ represents the total [CII] signal, $P_{\text{int}}(k, z)$ denotes the total interloper contamination, and $P_{\text{WN}}$ is the white noise power spectrum of the experiment. The error on the power spectrum is inversely proportional to the number of measured modes, $N_m$, which depends on the total survey volume, $V_{\text{surv}}$. We calculate $N_m$ using the prescriptions provided in \citep{limpy, Dumitru2018}.

\begin{figure*}
  \begin{center}
    \includegraphics[width=0.55\textwidth]{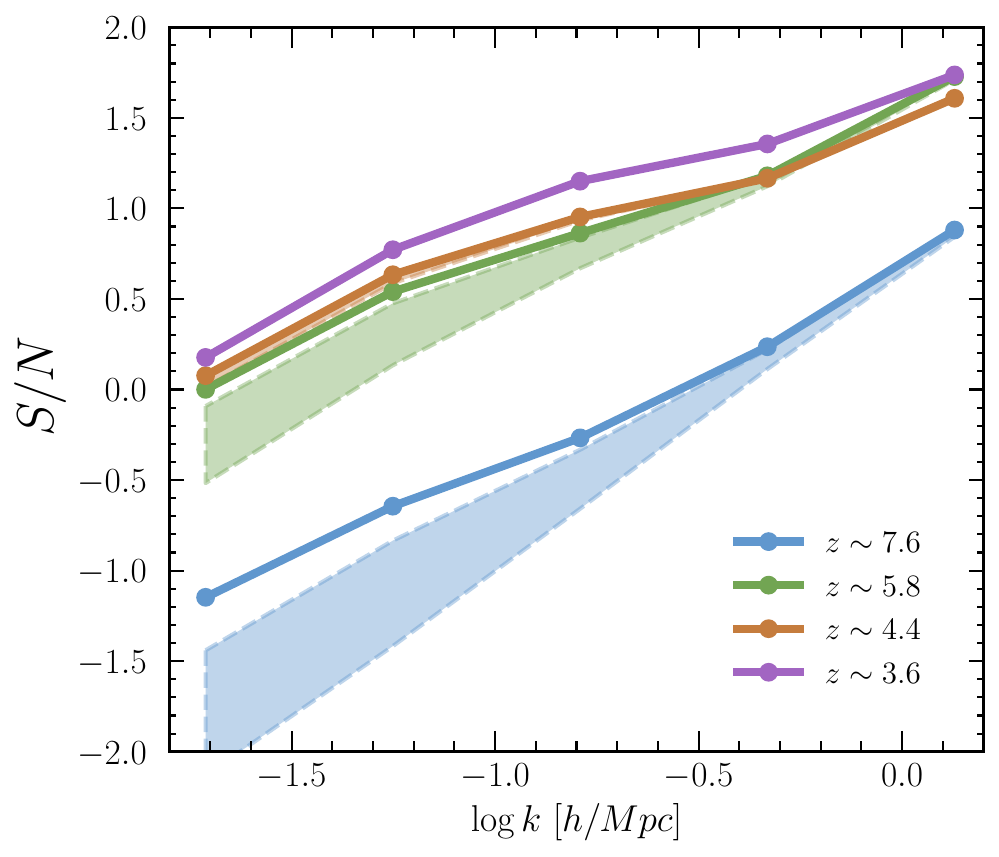}\\
    \caption{We forecast the signal-to-noise ratio (SNR) for detecting the [CII] signal using an instrument like EoR-Spec at FYST \cite{CCAT-prime2021}. The solid lines depict SNR values at four redshifts, distinguished by different colors in the legend, considering only instrument noise. The shaded regions of the same colors show SNRs when including both bright and faint interlopers along with instrument noise. Interlopers dominate at $z \sim 7.6$, reducing SNRs, whereas at $z \sim 3.6$, the signal dominates interlopers across all scales, with no impact on SNRs.}
    \label{fig|snr}
  \end{center}
\end{figure*}

\begin{table}
\centering
\renewcommand{\arraystretch}{1.3} 
\setlength{\tabcolsep}{6pt} 

\begin{tabular}{lccc|ccc}
\hline
\hline
  & \multicolumn{3}{c} {Noise} & \multicolumn{3}{c}{Noise/100} \\
\cline{2-4} \cline{4-7}
Redshift & Noise & Faint Int. & Bright Int. & Noise & Faint Int & Bright Int \\
\cline{1-7}
$z\sim 7.6$ & 13.6 & 13.2 & 11.7 & 889 & 501 & 119 \\

$z\sim 5.8$ & 105.7 & 103.2 & 94.5 & 1966 & 1772 & 998 \\

$z\sim 4.4$ & 89.5 & 89.3 & 87.9 & 1781 & 1763 & 1613 \\

$z\sim 3.6$ & 125.7 & 125.5 &  125 & 1837 & 1833 & 1797 \\

\cline{1-7}
\cline{1-7}
\end{tabular}
\caption{We present SNRs for detecting the [CII] signal in our model at redshifts $z \approx 7.6$, $5.8$, $4.4$, and $3.6$ under two scenarios. For one set of forecasted SNR, we consider instrumental noise to be the same as the noise of FYST's EoR-Spec-like experiment. In this case, the signal is more dominated by instrumental noise than interlopers. On the other hand, we forecast the SNRs by reducing the instrumental noise by two orders of magnitude so that we can understand the effect of SNRs primarily due to interlopers.}
\label{tab:mytable}
\end{table}

We present the signal-to-noise ratio (SNR) values for different \( k \)-bins at various redshifts of interest in Figure \ref{fig|snr} and summarize the SNR values for several scenarios in Table \ref{tab:mytable}. For an instrument like EoR-Spec at FYST, we forecast the SNR under two conditions: first, considering only instrumental noise as the source of contamination, and second, accounting for both instrumental noise and interlopers. We examine scenarios with both faint and bright interlopers. Our analysis reveals that the SNR is 13.6 when considering only instrumental noise. However, it decreases to 11.7 when both noise and bright interlopers (as modeled by \texttt{Kamenetzky15}) are taken into account. The impact of the bright interloper on the SNR is limited because the instrumental noise dominates both the interloper and the signal. Additionally, the SNR remains relatively stable when the target signal is much stronger than the interlopers. At $z \sim 3.6$, the SNR is approximately 125 across all scenarios. At $z\sim 5.8$, the SNR is 105 with only instrumental noise, 103 when including faint interlopers, and 94 with bright interlopers. Our forecast suggests that a FYST-like experiment will be able to detect the signal as predicted by our model with high confidence at the redshift range of 3.6 to 7.6 even in the presence of interlopers. Our forecast suggests that a FYST-like experiment will be able to detect the signal predicted by our model with high confidence across the redshift range of $3.6$ to $7.6$ even in the presence of interlopers.

Next, we consider an ideal scenario where instrumental noise is not the dominant factor, allowing us to clearly understand the influence of interlopers on the detection of the target signal. Therefore, we reduced the instrumental noise of EoR-Spec at FYST by a factor of 100. Under these conditions, we observe that the brightness of interlopers significantly affects the SNR, particularly at $z \sim 7.6$ and $z \sim 5.8$. At $z \sim 7.6$, the SNR reaches 889 when considering only instrumental noise. However, this value decreases substantially to 119 when accounting for bright interlopers along with noise. Similarly, at $z \sim 5.8$, the SNR is 1966 with only white noise, but it drops to 998 in the presence of bright interlopers and to 1772 with faint interlopers. Conversely, at $z \sim 3.6$, the target signal remains dominant over the interlopers across all scales. Here, the SNR is 1837 with only noise, and it slightly decreases to 1833 and 1797 when including faint and bright interlopers, respectively. 

\section{Discussion and Conclusion}\label{sec:discussion}

The study of multi-line intensity mapping offers the exciting prospect of detecting signals from the epoch of reionization and the post-reionization structure formation. It also provides a unique opportunity to probe the complex and intricate astrophysics of line emission across cosmic time. 
However, modeling intensity mapping from ab-initio approaches may be challenging, due to considerable astrophysical uncertainties and a noticeable degeneracy among astrophysical and cosmological parameters. In order to minimize these, we have developed a semi-empirical, data-driven framework of galaxy evolution, and we have exploited it to estimate the average intensity of the [CII] emission line across cosmic times. Our framework is based on the
observed galaxy stellar mass function, star-forming galaxy main sequence and fundamental metallicity relations. These quantities are then combined to stellar evolution and radiative transfer prescriptions of line emissions, to derive the cosmic [CII] intensity in the extended redshift range $z\sim 0-10$. The approach is fairly general and can be easily applied to other key lines for intensity mapping studies like [OIII] and the CO ladder. 

We forecast the detectability of the [CII] signal using an experiment analogous to FYST. In our analysis, we model the interlopers and categorize them based on their intensity, distinguishing between the brightest and faintest ones. We then evaluate the signal-to-noise ratio (SNR) across various redshifts to assess the impact of these interlopers on signal detection. Our findings indicate that interloper signals do not significantly affect the SNR for the detection of [CII] at redshifts below $z\approx 4.4$. However, their impact becomes pronounced if the instrumental noise is reduced to a level comparable to or lower than the interloper signals. In such cases, accurate modeling of interlopers is crucial for effectively extracting high-redshift signals. This underscores the importance of incorporating detailed interloper models in order to ensure reliable detection and analysis of signals from the galaxies in the early Universe.

In future work, we plan to extend our method to model additional atomic and molecular lines, such as CO(1-0) through CO(13-12), [OIII] at 88 and 52 $\mu$m, [CI] at 145 $\mu$m, among others. This extension will allow us to quantify both the target signal and the interlopers in a self-consistent manner. By predicting signals from multiple lines of interest, we will also be able to cross-correlate lines originating from the same redshift. This cross-correlation technique, as suggested by \cite{Roy-Lim-LLX}, has the potential to mitigate the effects of interlopers. Additionally, our semi-empirical framework provides a clear view of the environmental properties of the galaxies contributing to the line emission and thus can provide valuable insights into what astrophysics can be learned from LIM data. In this vein, we will explore the possibility of performing Fisher matrix analyses or Markov Chain Monte Carlo forecasts to constrain galaxy properties based on LIM observations. This approach will enhance our understanding of both the target signals and the underlying astrophysical processes.

\acknowledgments{AR thank Anthony Pullen and Nicholas Battaglia for useful discussions. AR acknowledges support from NASA under award number 80NSSC18K1014939. AL was partially funded from the projects: ``Data Science Methods for MultiMessenger Astrophysics \& Multi-Survey Cosmology'' funded by the Italian Ministry of University and Research, Programmazione triennale 2021/2023 (DM n.2503 dd. 9 December 2019), Programma Congiunto Scuole; EU H2020-MSCA-ITN-2019 n. 860744 \textit{BiD4BESt: Big Data applications for black hole Evolution STudies}; Italian Research Center on High Performance Computing Big Data and Quantum Computing (ICSC), project funded by European Union - NextGenerationEU - and National Recovery and Resilience Plan (NRRP) - Mission 4 Component 2 within the activities of Spoke 3 (Astrophysics and Cosmos Observations);  European Union - NextGenerationEU under the PRIN MUR 2022 project n. 20224JR28W "Charting unexplored avenues in Dark Matter"; INAF Large Grant 2022 funding scheme with the project "MeerKAT and LOFAR Team up: a Unique Radio Window on Galaxy/AGN co-Evolution; INAF GO-GTO Normal 2023 funding scheme with the project "Serendipitous H-ATLAS-fields Observations of Radio Extragalactic Sources (SHORES)".}

\bibliography{mybib}
\bibliographystyle{JHEP}
\end{document}